\documentclass{optica-article}

\journal{opticajournal}
\articletype{Research Article}

\usepackage[utf8]{inputenc}
\usepackage[T1]{fontenc}
\usepackage{siunitx}
\usepackage{dsfont}
\usepackage{eucal}
\usepackage[export]{adjustbox}
\usepackage{graphicx}
\usepackage{bm}
\usepackage{multirow}
\usepackage{tabularx}
\usepackage{dcolumn}
\usepackage{siunitx}

\begin{document}

\title{Evaluation of machine learning techniques for conditional generative adversarial networks in inverse design}

\author{Timo Gahlmann and Philippe Tassin}

\address{Department of Physics, Chalmers University of Technology, SE-41296 Göteborg, Sweden}

\begin{abstract*}
Recently, machine learning has been introduced in the inverse design of physical devices, i.e., the automatic generation of device geometries for a desired physical response. In particular, generative adversarial networks have been proposed as a promising approach for topological optimization, since such neural network models can perform free-form design and simultaneously take into account constraints imposed by the device's fabrication process. In this context, a plethora of techniques has been developed in the machine learning community. Here, we study to what extent new network architectures, such as dense residual networks, and other techniques like data augmentation, and the use of noise in the input channels of the discriminator can improve or speed up neural networks for inverse design of optical metasurfaces. We also investigate strategies for improving the convergence of the training of generative adversarial networks for inverse design, e.g., temporarily freezing the discriminator weights when the model outperforms the generator and training data blurring during the early epochs. Our results show that only some of these techniques improve inverse design models in terms of accuracy and stability, but also that a combination of them can provide more efficient and robust metasurface designs.
\end{abstract*}

\section{Introduction}

Neural networks are powerful machine learning tools that can learn complex patterns and functions from data~\cite{annbook2017,khan2018guide,Mehlig,goldberg2022nn}. They have been widely used for various tasks such as image recognition, natural language processing, and recommendation systems. However, their application to the inverse design of physical devices~\cite{Wiecha-review,Padilla2022} has been relatively limited. Inverse design~\cite{gahlmann2022deep,panisilvam2023asymmetric,Piggott,Michaels,Molesky,Whiting,P.Rodrigues,Kalt,Hegde,J.Mun,F.Cheng_2019,Alcalde,Pestourie,Piggott2020,X.Shi,Lulu_id,Kudyshev,Zhelyeznyakov,Zandehshahvar,PadillaLorentz2022,Psaltis2022,Adibi,Shin,Khoram} is the process of finding the optimal parameters or geometry of a device that can achieve a desired functionality or property, such as the optical response, thermal conductivity, or mechanical strength of a nanostructured material. This is a challenging problem because it normally involves solving nonlinear and high-dimensional optimization problems that are often constrained by physical laws and engineering requirements. Inverse design problems are often solved with optimization techniques such as binary search, physics-informed gradient-descent methods, and stochastic optimization like genetic algorithms and particle swarm optimization. However, these approaches require a large number of evaluations of the forward problem, solved sequentially, and can lead to capricious designs with small features that are difficult to fabricate.

\begin{figure}[htbp]
    \centering
    \includegraphics[width=0.83\textwidth]{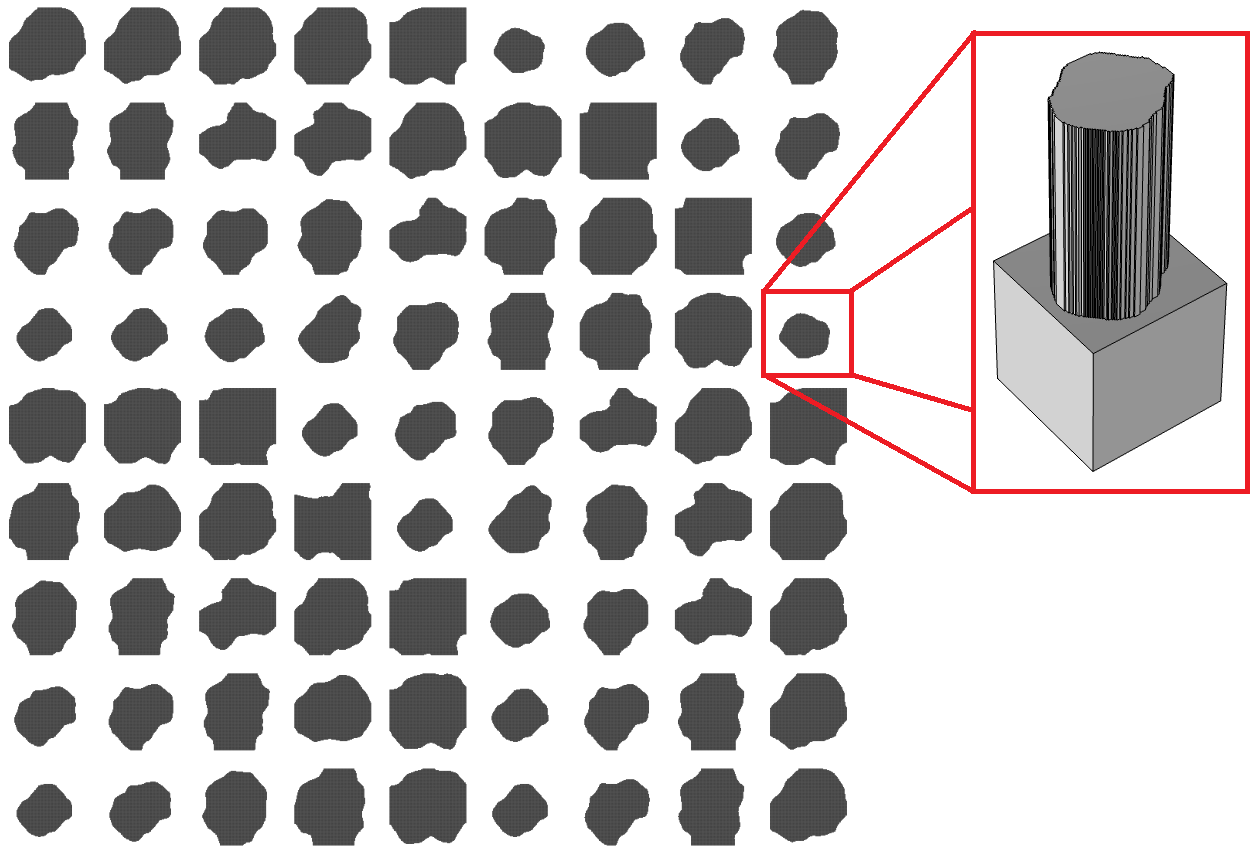}
    \caption{Top view of a metasurface, consisting of a grid of pillars etched out of a substrate. The figure represents the etching mask used in the metasurface's fabrication, creating pillars on a substrate with the same cross-section (see inset).}
    \label{Fig:metasurface}
\end{figure}

One way to overcome this difficulty is to use deep-neural-network (DNN) models to achieve inverse design of physical devices. This is often implemented using a forward network that predicts physical properties for a device with arbitrary geometry, and then an inverse model is constructed using a tandem network~\cite{Khoram} or a conditional generative-adversarial network (CGAN)~\cite{Goodfellow_2014,mirza2014conditional,hong2018conditional,B.Zheng,J.Jiang,S.SoandJ.Rho}. The key advantage of DNNs in the design of physical devices is their ability to quickly generate designs for devices with any desired, physically allowed response. For example, in an optical metasurface we can design meta-atoms with any desired transmission amplitude and phase in the subset of realizable S parameters once trained, eliminating the need to run an optimization for every and each meta-atom. The training of the neural networks still requires a large number of simulations of the forward problem, but these simulations are independent, meaning they can be run simultaneously on high-performance computer clusters.

\begin{figure*}[htbp]
    \centering
    \includegraphics[width=1.0\textwidth]{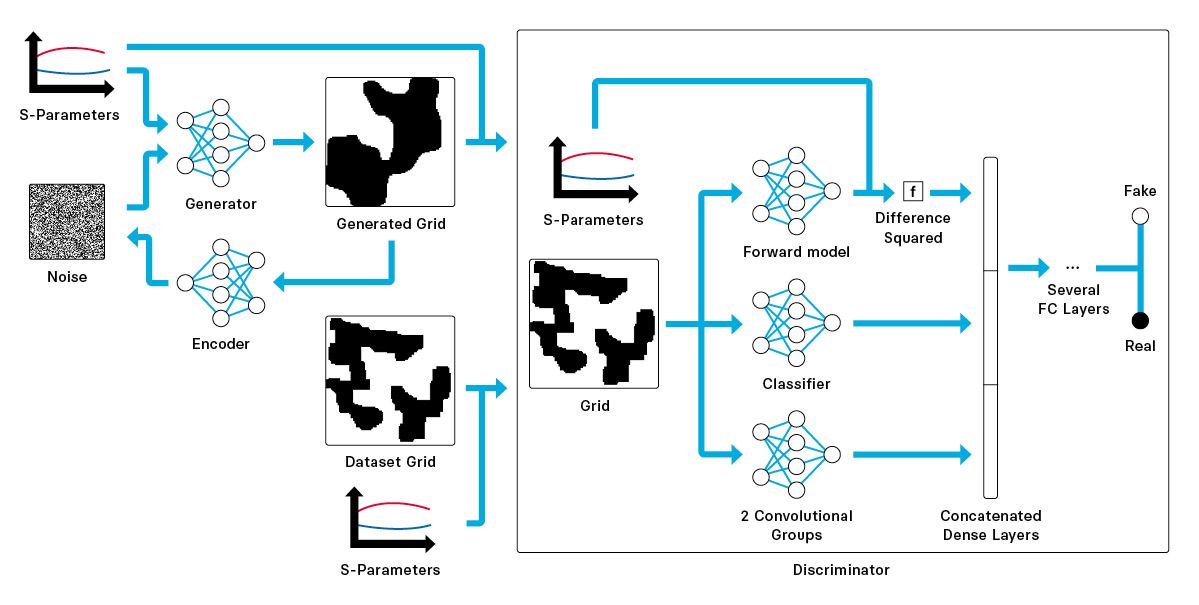}
    \caption{Schematic of the conditional generative adversarial network model and its components. The generator creates meta-atom grids from noise and $\mathcal{S}$ parameters, while the encoder prevents mode collapse. Two expert networks (forward and classifier model) are integrated inside the discriminator. Reprinted from~\cite{gahlmann2022deep}.}
    \label{Fig:cGAN}
\end{figure*}

In this article, we study a number of different machine-learning techniques for enhancing conditional generative adversarial networks (CGANs) for the inverse design of optical metasurfaces. Such nanophotonic structures are thin optical structures, often consisting of a patterned structure on a substrate (see Fig.~\ref{Fig:metasurface}). Each element (see inset), also called meta-atom, can locally change the amplitude and phase of the transmitted and reflected electromagnetically waves. When fabricating such optical metasurfaces, the shape of the elements is etched out of a substrate using a lithographic mask, similar to what is shown in Fig.~\ref{Fig:metasurface}. By creating an array of different meta-atoms, we can imprint a transmission or reflection mask, e.g., mimicking the function of a lens or any other optical device.

A schematic of the CGAN we are using is shown in figure~\ref{Fig:cGAN}. Initially, a forward model is established to approximate the optical properties of meta-atoms (here transmission or reflection) as a function of the grid representing the lithographic mask. This forward model accelerates the assessment of optical properties compared to direct full-wave solutions. A binary grid is utilized to represent the lithographic mask of the meta-atom, with 0 indicating that the top layer is etched away and 1 meaning the top layer remains unetched. Due to similarities with pixelated images, a convolutional neural network (CNN) in the forward model is used to model the scattering parameters. This CNN is trained with about 40,000-80,000 semirandom grids created and loaded into a finite-element simulator to obtain the scattering parameters. 

The binary grids are generated from random noise by filling small holes and erasing small islands of increasing size, as such features cannot be realized by current lithographic techniques. Hence, we use a free-form representation of metasurface designs not restricted to specific geometries such as cylinders or cubes. In this context, an important aspect is the integration of fabrication feasibility into the inverse design method. This issue is addressed by training a classifier network on flawless and flawed grids. The grids are labeled with the number of pixels that would need to be changed to satisfy the fabrication restrictions and the classifier network is trained to approximate these. The issue of the ``one-to-many'' problem~\cite{hadamard1902problemes} in inverse design, i.e., the fact that multiple non-unique solutions exist for a given design target, is solved using a conditional generative adversarial network (CGAN). The CGAN is designed to generate grids based on noise, desired scattering parameters, and fabrication constraints. The discriminator of the CGAN, which discerns between real and generated data, also evaluates whether the grids have the desired optical properties and comply with the required fabrication constraints. Furthermore, an encoder network helps with solving the issue of mode collapse (where the generator maps only to a small subset of the model) and enables a greater variety of generated grids. This combination of DNNs and our enhanced CGAN approach offers a powerful and efficient solution to the complex problem of metasurface design, paving the way for advancements in the field of nanophotonic structures.

In this neural-network model for inverse design, we then study the utility of integrating the forward model and the classifier model into the discriminator, the use of modern network architectures based on dense residual networks compared to simpler convolutional networks, the use of data augmentation, and the use of noise in the input channels of the discriminator. We also investigate a strategy for training the discriminator on other sources of generator outputs and temporarily freezing the discriminator weights to improve stability during training. We study whether these techniques improve inverse design in terms of relevant metrics like validation errors or S-parameter accuracy in order to quantify their effectiveness. Our results show that a combination of these techniques outperforms previous methods in terms of accuracy, stability, and efficiency while providing more robust metasurface designs, but also that not all of the above-mentioned machine-learning techniques improve inverse design.

\section{Machine-learning techniques}

\subsection{Network architecture}

In recent years, neural network architectures have been the subject of intense research and development. Convolutional neural networks (CNNs) have become the architecture of choice for image-related tasks. However, standard CNNs have a number of limitations, such as vanishing gradients, which can make it difficult for them to learn from deeper layers in the network. To address these issues, researchers have developed various architectures, such as residual networks (ResNets)~\cite{ResNet}, which use skip connections to enable information to flow more freely through the network. One of the newer developments in neural-network architectures is dense residual networks (DenseNets)~\cite{huang2017densely}, which take the idea of skip connections a step further by connecting a lot of layers directly with each other. This results in a very dense and connected network structure that allows for efficient feature propagation and reuse. The DenseNet architecture has shown great promise in various image-related tasks, including object detection and image segmentation~\cite{huang2018condensenet,iandola2014densenet,zhang2019multiple}.

In this paper, we study whether DenseNets can improve learning efficiency from the available training data in CGAN network architectures for inverse design. The significance of DenseNets lies in their ability to alleviate the limitations of traditional CNNs and ResNets, leading to improved performance and faster training times. We will use DenseNets in the classifier and forward models, as well as in the discriminator of the CGAN. The full network architecture of the different parts of the CGAN can be found in the supplementary material and will be referenced when discussed in the text below.

\subsection{Data augmentation}\label{Chap:aug}

Data augmentation~\cite{mikolajczyk2018data,khalifa2022comprehensive,chlap2021review,kostrikov2020image} is a technique that has become increasingly popular in machine learning, and it is particularly relevant in the context of training neural networks. The idea behind data augmentation is to artificially increase the amount of training data by generating new examples that are similar to the original ones but contain variations in some aspects. This can be done by applying various transformations to the original data, such as rotations, translations, scaling, flipping, or adding noise. 

The primary reason for using data augmentation is to improve the performance and robustness of the neural network. By training on a larger and more diverse set of data, the network can learn to generalize better and be more capable of handling variations and noise in the real-world data. Additionally, data augmentation can help to reduce overfitting. Furthermore, data augmentation is particularly important in domains where the amount of available data is limited, such as in scientific or medical applications. In these cases, it is often not feasible to collect large amounts of new data, and therefore data augmentation can be an effective way to leverage the existing data and improve the performance of the network. For inverse design, data augmentation can reduce the need for training data, which is important as the generation of training data through simulations constitutes the major computational cost.

\begin{figure}[htbp]
    \centering
    \includegraphics[width=0.55\textwidth]{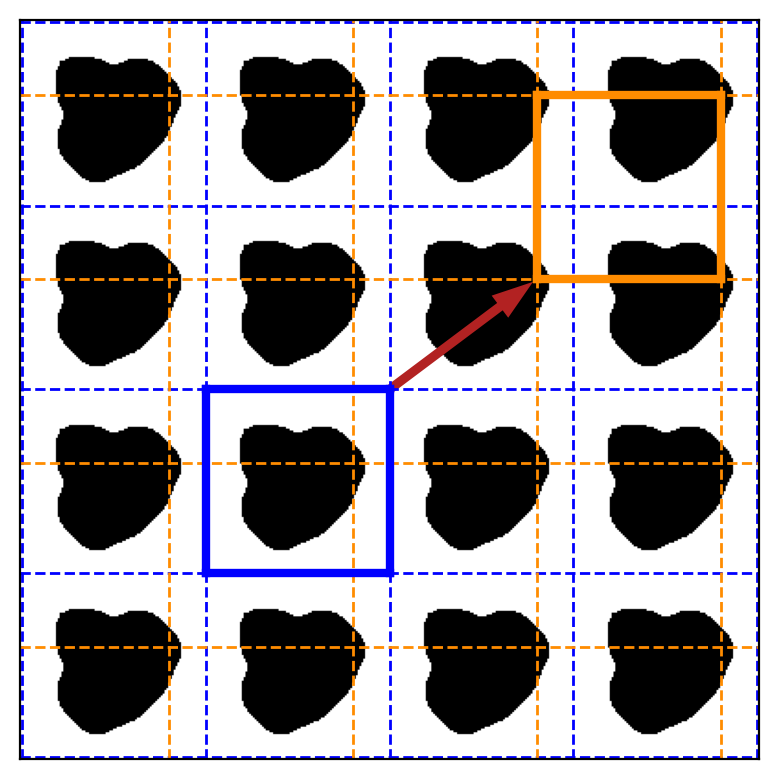}
    \caption{Visualization of translational augmentation.}
    \label{Fig:aug_vis}
\end{figure}

For the metasurfaces we are working with in this article, we cannot apply arbitrary rotation or scaling to the training data, since these operations on the unit cell alter the optical properties. However, translation can be used because the unit-cell simulations we use as training data have periodic boundary conditions, essentially creating an infinitely repeating surface. Therefore, a translation by $n\cdot\frac{\vec{a}}{p_a}$ or $n\cdot\frac{\vec{b}}{p_b}$, where $p_{a,b}$ denote the pixel count in the respective dimension and $n \in \mathds{N}$, provides an equivalent representation of the unit cell with the same optical response. This concept is illustrated in figure~\ref{Fig:aug_vis}. In addition to translation, $90^\circ$, $180^\circ$, and $270^\circ$ rotations of the unit cell can also be employed in the classifier training, since the experimental feasibility is not affected by this transformation, and could be employed as well in the forward model training by making an appropriate transformation of the S parameters.

\subsection{Noise}
Just like using dropout, introducing noise~\cite{piotrowski2013comparisonnoise,zur2009noise,noh2017regularizing} into the networks during training is a technique that can be used to avoid overfitting and increase robustness. Here, we use blurring~\cite{wang2019blur,vasiljevic2016examining_blur} of the training data and Gaussian noise to optimize the convergence and performance of our networks. We apply noise with a larger standard deviation in the forward and classifier models' input channels compared to the discriminator's own convolution input channel (starting at 1.1 and 0.3, respectively) and decreasing over the first 15 epochs. Different kinds of blurring were employed by applying semirandom convolutions to random subsets of the training data. These semirandom convolutions were generated by randomly picking an envelope function (Gaussian: $b_g(r)=\exp\left(-r^2/2 a_g^2\right)$ with $a_g\in [0.5,4]$; linear: $b_l(r)=1-r/a_l$ for $b_l(r)>0$ with $a_l \in [2,8]$ and $b_l(r)=0$ otherwise; or inverse: $b_i(r)=\frac{1}{r+1}$), and a convolution size/range between 3 times 3 and maximally 15 times 15 (the maximum is decreasing with each epoch). Then, for each pixel in the convolution, the envelope function (where $r$ is the distance of the pixel from the center) is multiplied with a random number between 0 and 1 and the sum of the pixel values in the convolution is normalized to 1.

\section{Results for the classifier and forward networks}
In this section, we analyze the effects that the different techniques had on the training of the classifier and forward networks. To this end, we plot loss functions of the individual training runs for their respective epoch lengths from which we can compare the performance of two groups, one group of curves where a particular technique (red curves) is used and another group where the technique is not used (blue curves). To analyze the statistics of these curves, we calculated the p-values for each of the groups with the two-sided Kolmogorov-Smirnov test, where the null hypothesis is that there is no difference between the two groups and we evaluate the Kolmogorov-Smirnov test on the basis of the minima of the validation loss of each training run, since these represent the quality of the saved models.

\textbf{DenseNets} We first investigate the effect of using DenseNets in the forward network, mapping the geometry of the metasurface on the optical properties, and in the classifier network. The full networks of the forward network with DenseNet and without DenseNet can be found in figures~\ref{Fig:network_forward_withdensenet} and \ref{Fig:network_forward_withoutdensenet}, respectively, in the supplementary material.

The full networks for the classifier can be found in figures~\ref{Fig:network_classifier_withdensenet} and \ref{Fig:network_classifier_withoutdensenet}.

\begin{figure*}[htbp]
    \centering
    \includegraphics[width=0.50\textwidth]{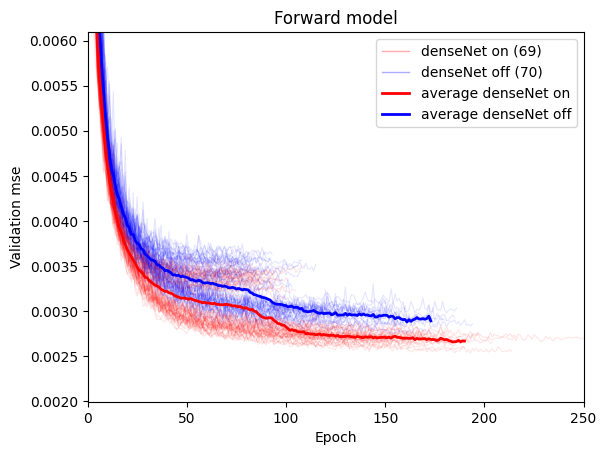}
    \includegraphics[width=0.470\textwidth]{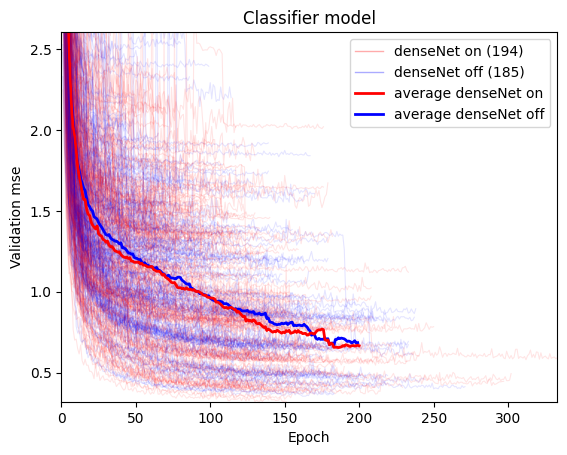}
    \caption{Comparison of the validation loss in the forward and classifier models between models with a DenseNet architecture and models with a standard 2D convolution. The numbers in the legend indicate how many models were trained in each category. P-values: forward model $p<10^{-5}$, classifier model $p=0.158$.}
    \label{Fig:DenseNet_comparison}
\end{figure*}

In figure~\ref{Fig:DenseNet_comparison}, we plot the mean square error for the validation sets for the forward model and for the classifier model with and without the DenseNet architecture.
Each line in the graph represents a DNN trained with 50,000 training samples randomly selected from the full data set (roughly 80,000 samples). Red lines signify models with DenseNets and blue lines models with a generic 2D convolutional network. The thick lines represent an average over the different models. We clearly see that the adoption of the DenseNet architecture improves the forward networks (the average validation loss line with a DenseNet lies clearly lower and there is a very low probability that the null hypothesis is valid (no difference between the red and blue groups) is valid: $p<10^{-5}$). For the forward network, we see two clusters of lines, which is because augmentation (see below) is an even more important contributor to the training and divides the samples into said clusters. The DenseNets make up the lower part of each cluster. For the classifier network, the spread for different networks is larger, but we see that the best networks use the DenseNet architecture and also that on average it is better to use DenseNets in the forward network. This trend is not reflected in the classifier training runs, where it makes little to no difference whether a DenseNet or a generic 2D convolutional network is used (the average validation loss lines for a classifier model with and without DenseNets coincide; the p-value is $p=0.158$).

\begin{figure*}[h]
    \centering
    \includegraphics[width=0.5\textwidth]{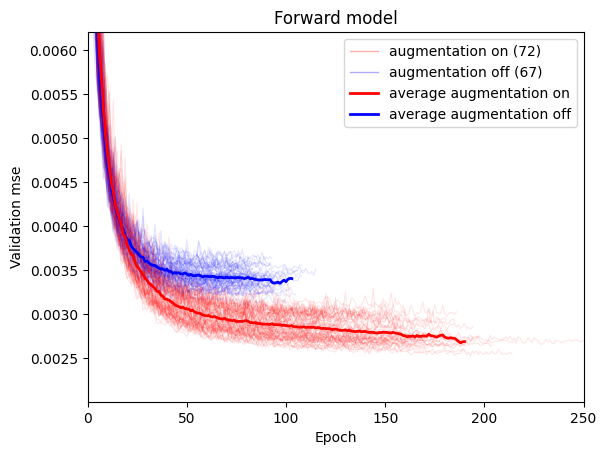}
    \includegraphics[width=0.470\textwidth]{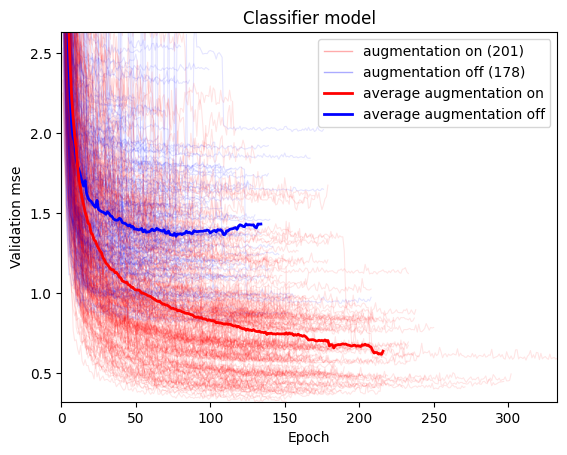}
    \caption{Comparison of the validation loss in the forward and classifier model with and without augmentation. The numbers in the legend indicate how many models were trained in each category. P-values: forward model $p<10^{-5}$, classifier model $p<10^{-5}$.}
    \label{Fig:Augmentation_comparison}
\end{figure*}

\textbf{Data augmentation} Figure~\ref{Fig:Augmentation_comparison} is a plot of the same data set as presented in figure~\ref{Fig:DenseNet_comparison}, but now the data is organized according to whether data augmentation has been utilized. Red lines are models that take advantage of data augmentation; blue lines represent models without data augmentation. From this figure it is evident that data augmentation exerts a considerably more substantial influence on the validation errors of the two networks compared to the DenseNet architecture. The two clusters that we saw before in the data for the forward model are clearly related to the use of data augmentation, which is reflected in the p-values (forward model: $p<10^{-5}$, classifier model $p<10^{-5}$).

\begin{figure*}[h]
    \centering
    \includegraphics[width=0.5\textwidth]{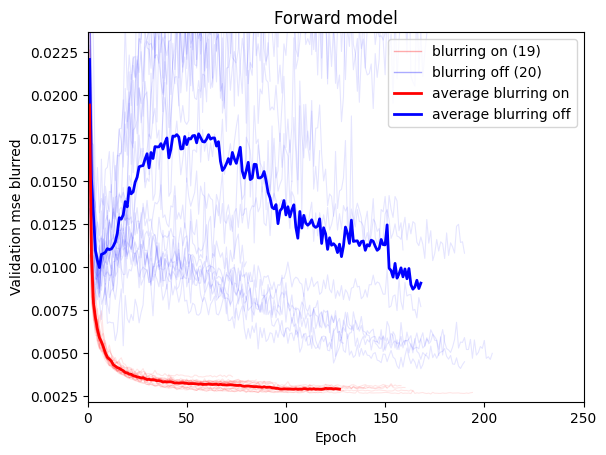}
    \includegraphics[width=0.465\textwidth]{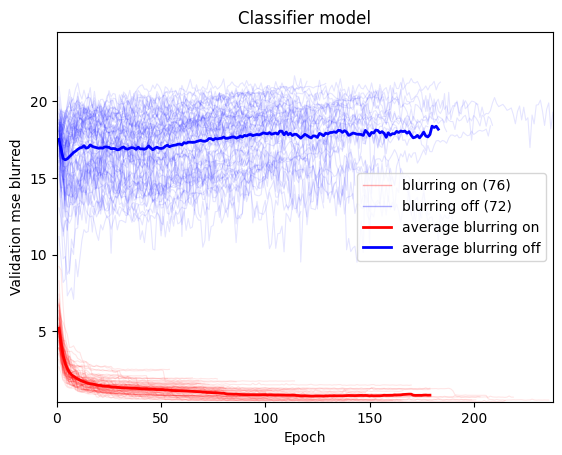}
    \caption{Comparison of the validation loss (blurred) in the forward and classifier model with and without blurring of the training data. The numbers in the legend indicate how many models were trained in each category. P-values: forward model $p<10^{-5}$, classifier model $p<10^{-5}$.}
    \label{Fig:blurring_comparison}
\end{figure*}

\textbf{Blurring} Finally, we look at the effect of blurring of the grids on the classifier and forward models. This is especially important for inverse design with CGANs, because it will enhance their accuracy when presented with outputs of the generator, which often resemble blurred versions of the training data grids, especially in the initial phase of the training. Therefore, we compare the performance of networks trained on partially blurred data to networks trained on data that was not blurred. As shown in figure~\ref{Fig:blurring_comparison}, the effect of blurring parts of the training data is significant, since it substantially improves the accuracy of the networks when they are presented with blurred unit cells---indeed, both the forward model and the classifier model have a p-value below $10^{-5}$. This does not affect the performance of the networks when presented with nonblurred data in a relevant way. Consequently, in the initial and intermediate stages of training, the CGAN discriminator receives more accurate information from both the forward and classifier models about the generated unit cells.

\section{Results for the CGAN}
In this section, we analyze the effects that the different techniques had on the training of the CGAN networks. To this end, we plot the $\mathcal{S}$ parameter MSEs, classifier outputs, and pixel errors of the generated unit cells for the individual training runs for 150 epochs, again comparing a group of training runs where a specific technique is used (red lines) to a group where the technique is not used (blue lines). To analyze the statistics, we calculated the p-values of all three metrics for each of the two groups with the two-sided Kolmogorov-Smirnov test. The null hypothesis is that there is no difference between the two groups and we evaluate the Kolmogorov-Smirnov test on the basis of the average value of the individual metrics of the epochs 120 to 150.

\subsection{Discriminator}
The success of CGANs~\cite{Goodfellow_2014,mirza2014conditional,hong2018conditional} is highly dependent on the quality of their discriminator, which acts as the dynamic error function of the generator. The discriminator is responsible for evaluating the quality of the generated meta-atoms by comparing them to (i.e., meta-atoms from the training data set). It is trained to distinguish between the real and generated meta-atoms, and in doing so, provides feedback to the generator on how to improve its output, which in turn is trained to fool the discriminator into believing that its output is part of the training data. Therefore, the discriminator plays a critical role in ensuring that the generated meta-atoms are of high quality and meet the desired specifications. However, setting up the training loop in a way that promotes the training of a discriminator that is punishing the generator for all relevant flaws can be a challenge. If the discriminator is too strong in identifying a single aspect of the generator output flaws, the generator may improve on this single aspect of the meta-atoms only, and neglect the other relevant criteria. Other issues arise if the discriminator is too weak and therefore not able to accurately distinguish between real and generated meta-atoms, which almost always results in poor quality outputs. This is one of the major challenges in CGAN-based inverse design. Here we evaluate the performance of the generator based on the S-parameter accuracy, the classifier output of the generated meta atoms, and the pixel error, which is a measure of how blurry the outputs are. The pixel error is defined as a sum over all pixels $\sum_i x_i^2(x_i-1)^2$, where $x_i$ is the individual value of each pixel. We explored different techniques for improving the stability of the discriminator relative to the generator. The most important parameters in this context are the optimizer and the learning rates. For the generator we used the Adamax optimizer with learning rate $\in [5\cdot 10^{-5},2\cdot 10^{-4}], \beta_1=0.9, \beta_2=0.999, \epsilon=10^{-7}$ and for the discriminator Adam with learning rate $\in [5\cdot 10^{-8},8\cdot 10^{-7}], \beta_1=0.333, \beta_2=0.999, \epsilon=10^{-7}$. These parameters are chosen such that the discriminator generally has a slight advantage over the generator in the long term.
\begin{figure*}[h]
    \centering
    \includegraphics[width=0.45\textwidth]{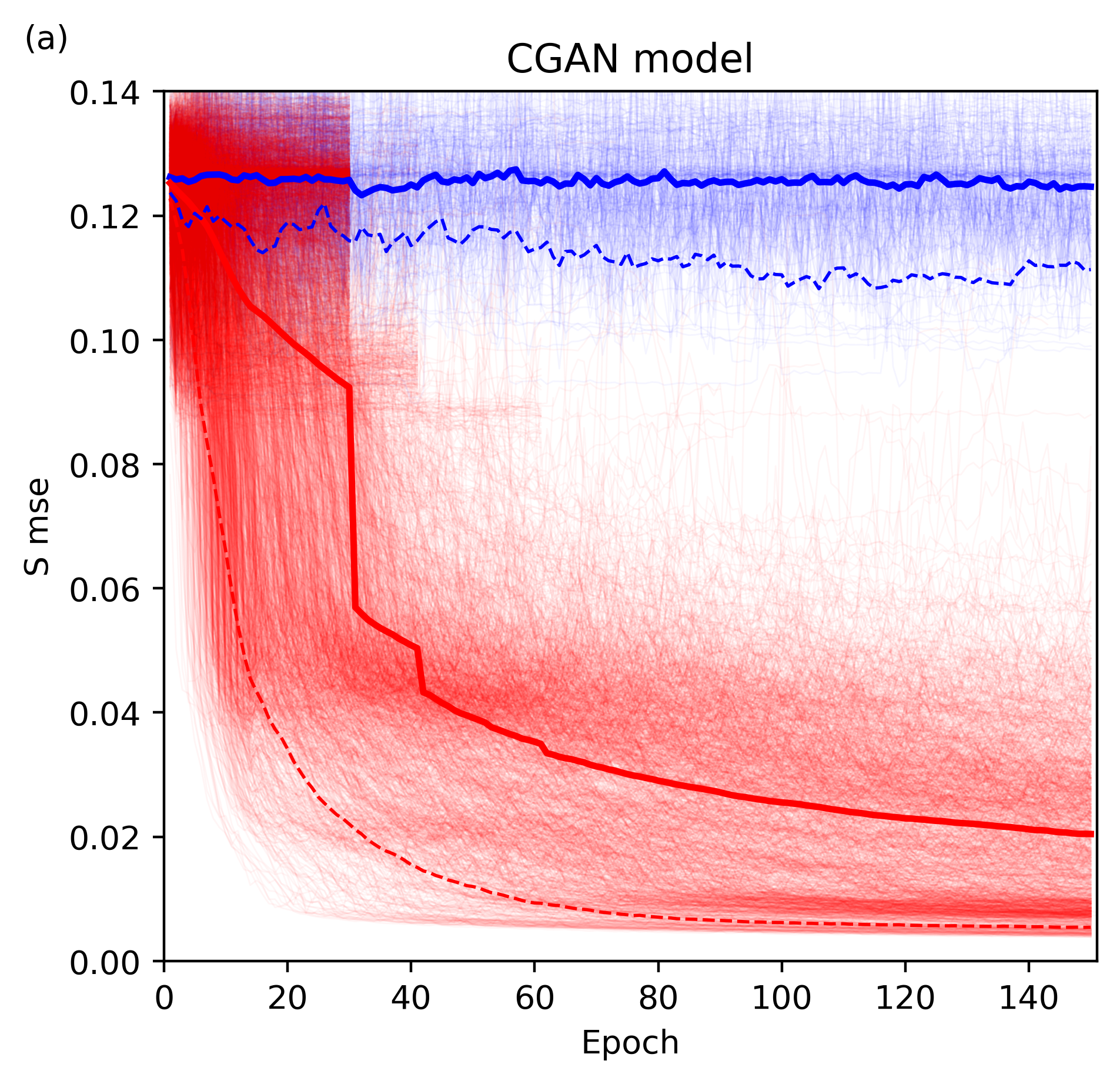}
    \includegraphics[width=0.45\textwidth]{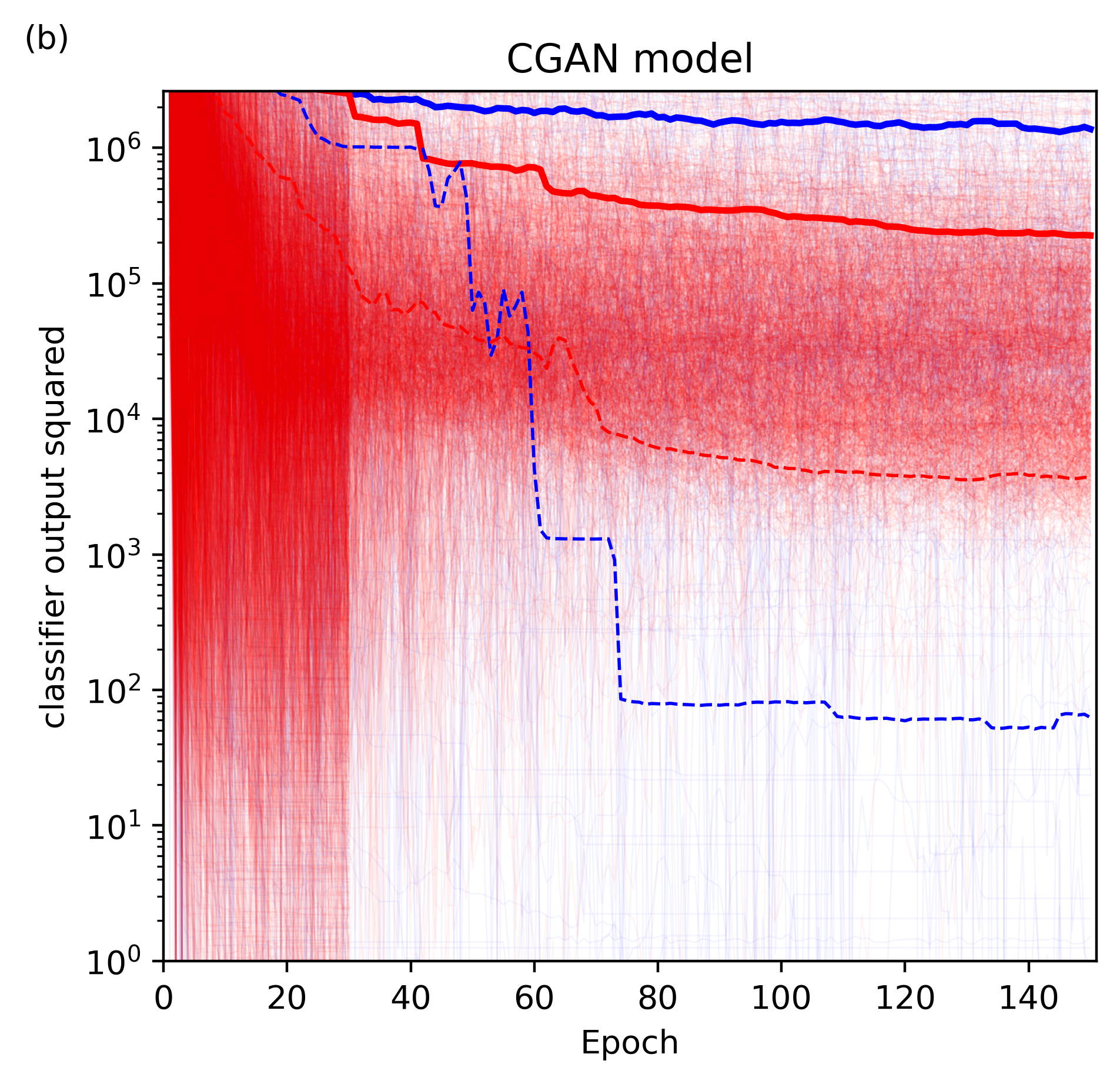}
    \includegraphics[width=0.45\textwidth,valign=m]{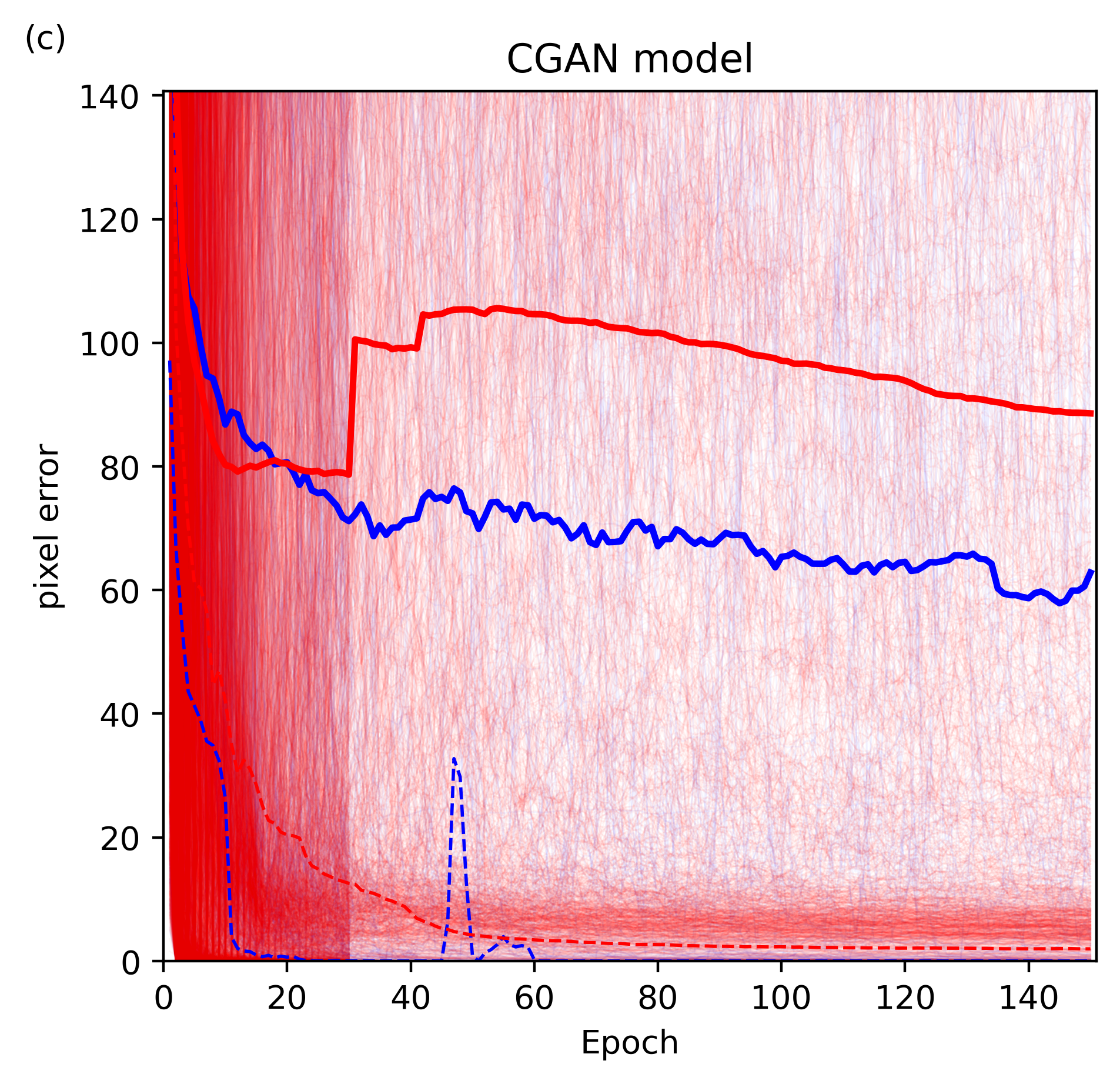}
    \includegraphics[width=0.45\textwidth,valign=m]{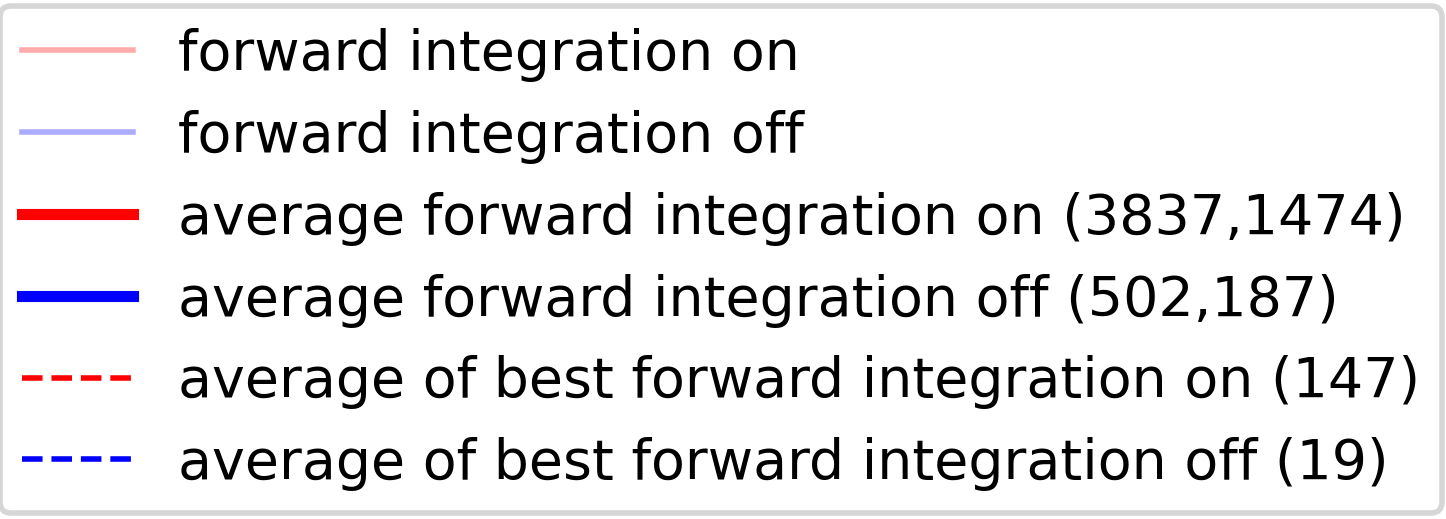}
    \caption{Comparison of the effects that the integration of the forward network in the Discriminator has on (a) $\mathcal{S}$ parameter MSE, $p<10^{-5}$, (b) classifier output, $p<10^{-5}$, and (c) pixel error, $p<10^{-5}$. In the legend, numbers after overall averages denote the total number of training instances and those reaching epoch 150. Similarly, the figure following the best training instance average indicates the number of instances used for that average.}
    \label{Fig:cgan_fwdInDisc_comparison_cgan}
\end{figure*}

\begin{figure*}[h]
    \centering
    \includegraphics[width=0.45\textwidth]{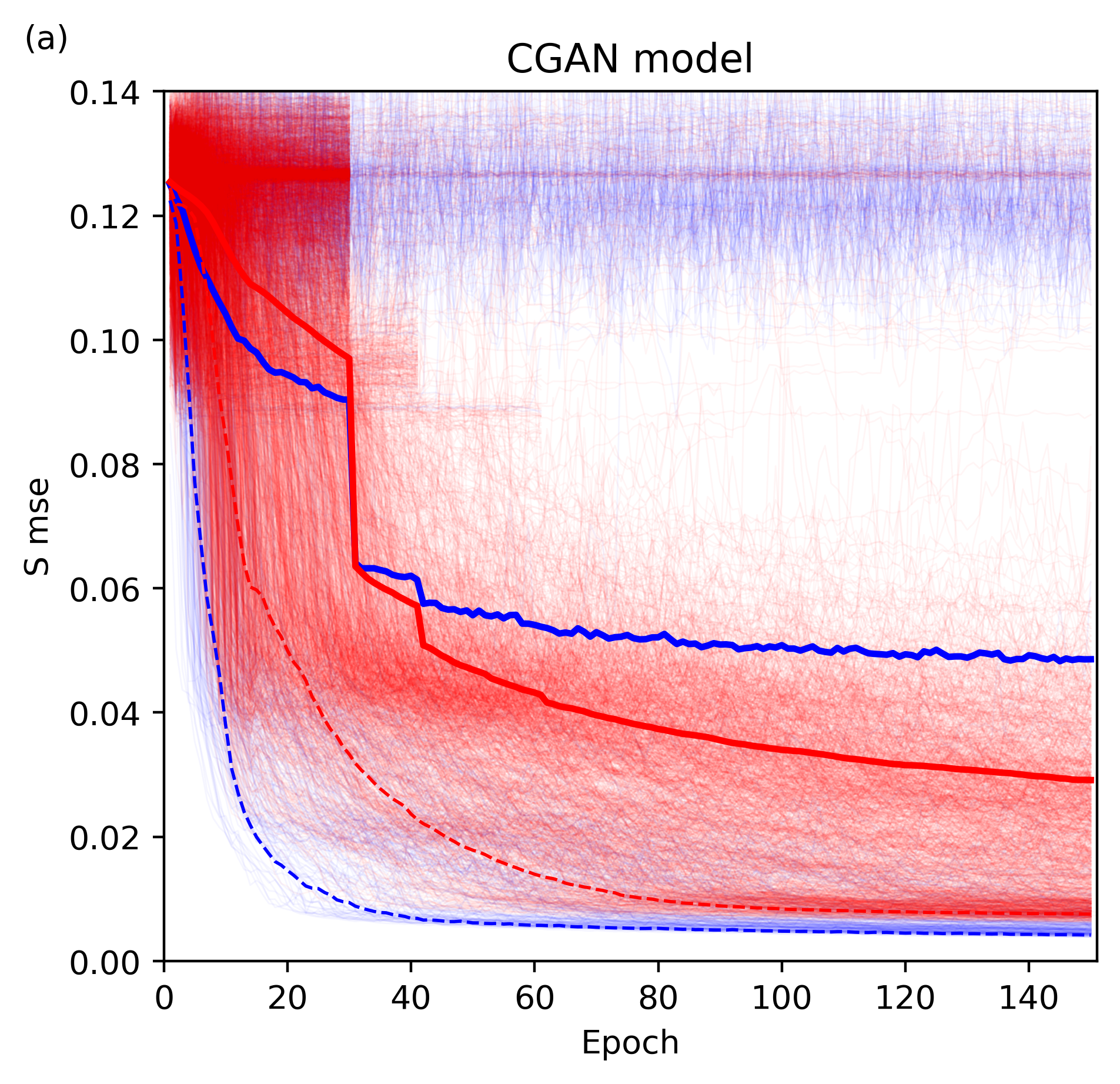}
    \includegraphics[width=0.45\textwidth]{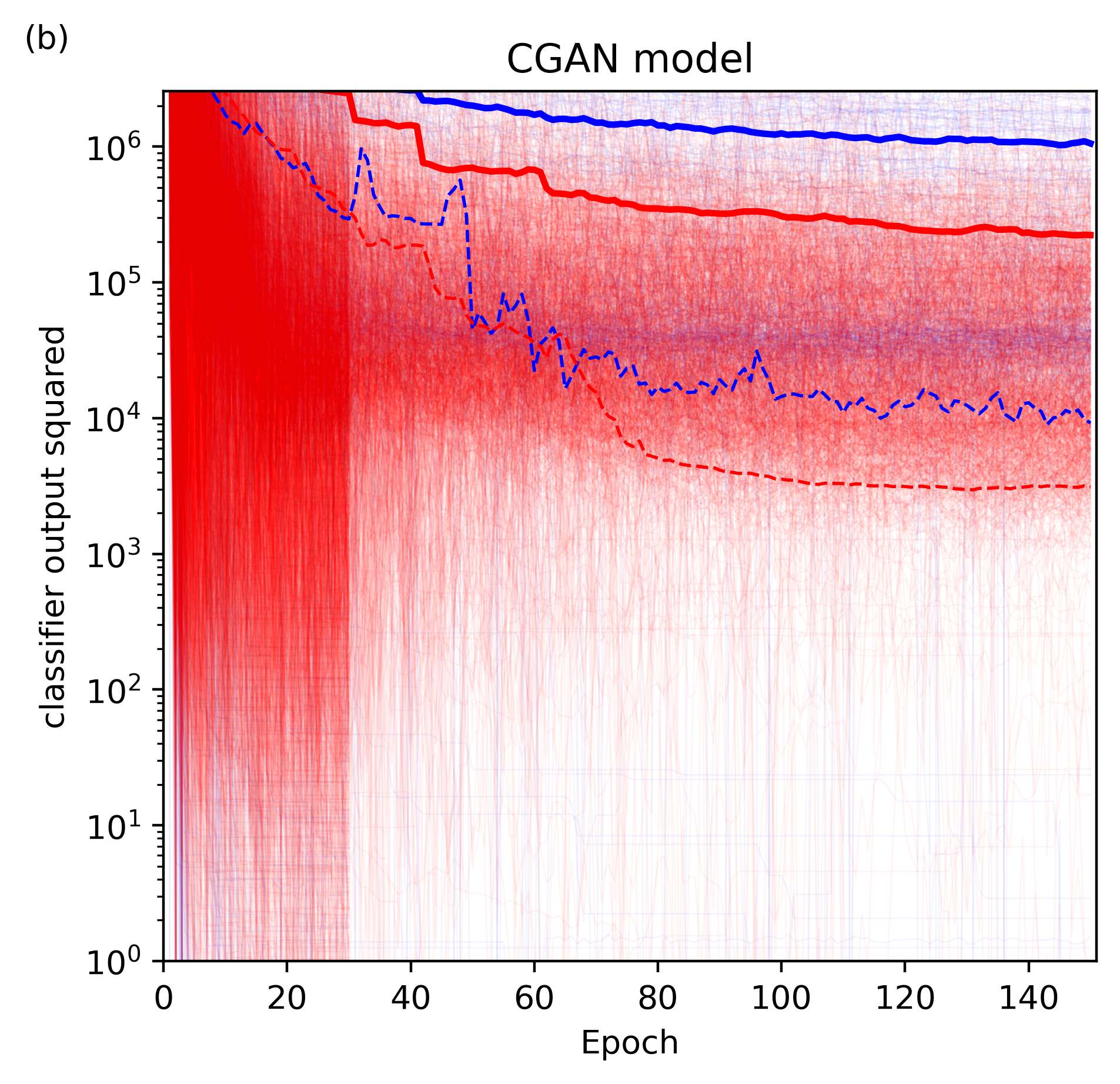}
    \includegraphics[width=0.45\textwidth,valign=m]{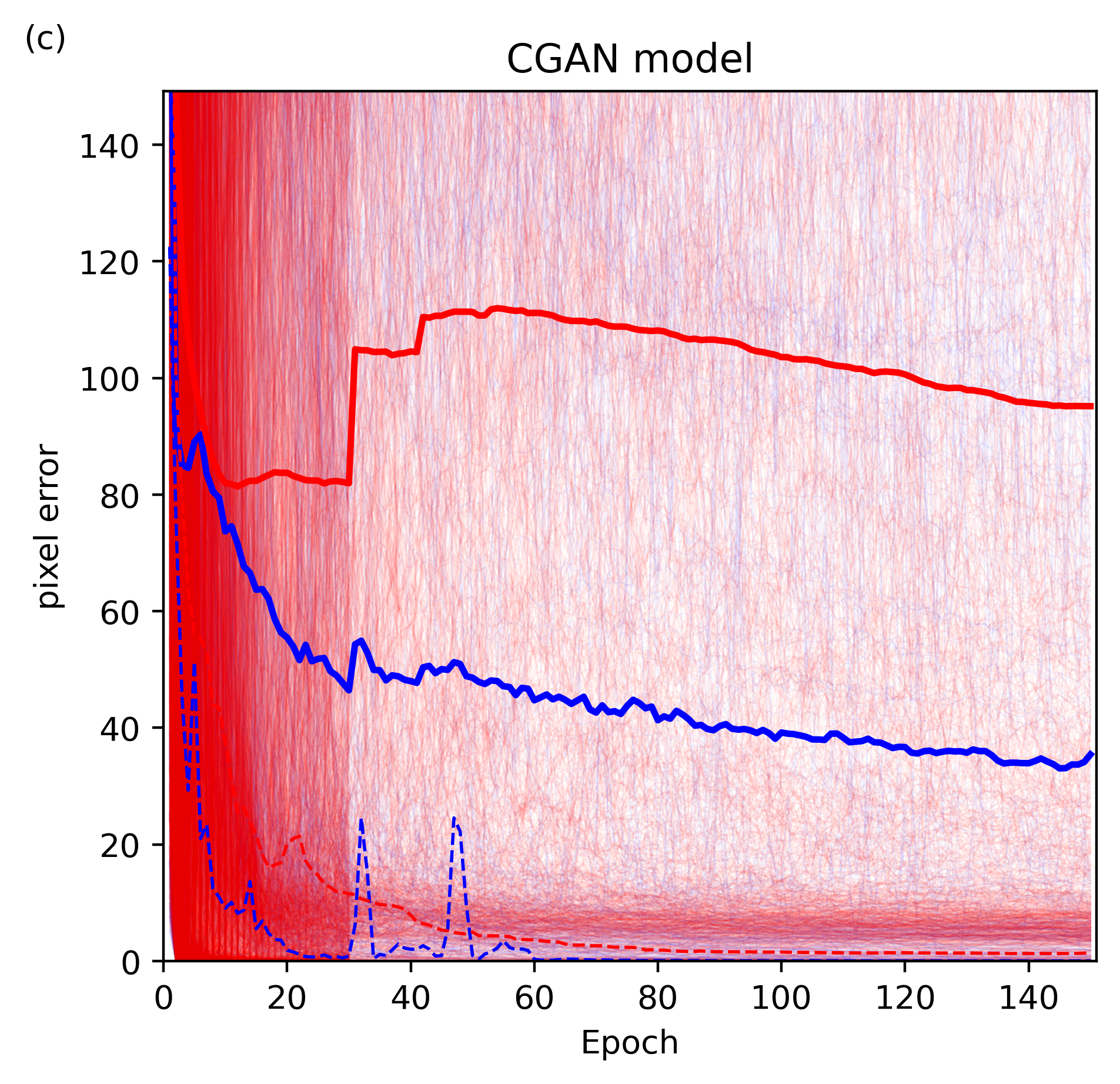}
    \includegraphics[width=0.45\textwidth,valign=m]{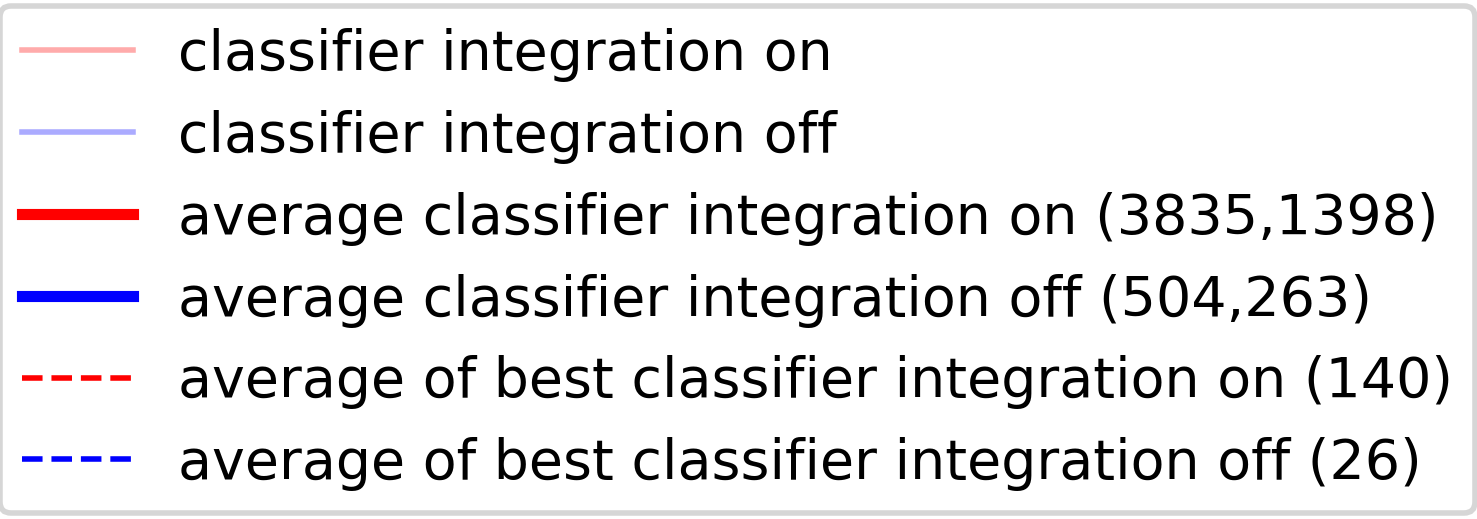}
    \caption{Comparison of the effects that the integration of the experimental feasibility classifier network in the Discriminator has on (a) $\mathcal{S}$ parameter MSE, $p<10^{-5}$, (b) classifier output, $p<10^{-5}$, and (c) pixel error, $p<10^{-5}$. In the legend, numbers after overall averages denote the total number of training instances and those reaching epoch 150. Similarly, the figure following the best training instance average indicates the number of instances used for that average.}
    \label{Fig:cgan_clInDisc_comparison_cgan}
\end{figure*}

\textbf{Expert model integration} As explained in the introduction, we integrated the forward model and the experimental-feasibility classifier in the discriminator of the CGAN in order to improve the discriminator's and thereby the generator's performance. The comparisons in performance can be assessed in figure \ref{Fig:cgan_fwdInDisc_comparison_cgan} and \ref{Fig:cgan_clInDisc_comparison_cgan}, respectively. With the integration of the (pre-trained) forward model enabled, the S-parameter error decreases significantly during the training, while it does not converge with the integration of the forward model disabled ($p<10^{-5}$). This shows that the forward model in the discriminator is essential for CGAN-based inverse design. Disabling the classifier integration in the discriminator increases the S-parameter accuracy of the generator outputs, since the generator is getting significantly less punished for unfeasible grid designs and, therefore, has greater freedom to find lithographic masks that possesses the desired optical properties. However, as can be seen from the classifier outputs, these designs tend to be unmanufacturable, as shown in figure~\ref{Fig:cl_in_disc_comp}. Note that the masks generated with the classifier integrated are smooth, while the masks generated without the classifier integrated contain very small islands, have rough edges, and narrow pathways between large islands—all features that cannot be manufactured. Similarly to the forward network, the classifier is essential in achieving masks that satisfy the fabrication constraints.

The parts of the data set where either the integration of the forward or the classifier or both were disabled are excluded from the comparisons below, since some of the conditions for early training terminations were disabled and those modifications alter the behavior of the discriminator too drastically and would therefore pollute the other data.

\textbf{DenseNets} To enhance the discriminator's analytical capabilities and improve its ability to assess training data and generator outputs while identifying discrepancies, a DenseNet~\cite{huang2017densely} was incorporated into the discriminator (network architectures can be found in figure~\ref{Fig:discriminator_DenseNet} and~\ref{Fig:discriminator_conv} in the supplementary material). The effects of the DenseNet compared to a standard convolutional network can be seen in figure \ref{Fig:cgan_denseNet_comparison_cgan} (see supplementary material). The improvements that we expected did not materialize. It is not entirely clear to us why that is the case.

\textbf{Discriminator freezing} In an attempt to balance the advantage of the discriminator in the long term, we tried skipping the discriminator training in each batch with a probability proportional to the generator loss in the previous batch. However, we never allow the probability of skipping discriminator training to reach one, as this results in a static discriminator and consequently a static error landscape of the generator, which can ultimately lead to a stagnation of the CGAN learning process. Adaptive skipping of the discriminator training proved to be essential in stabilizing the loss function ratio of the generator and discriminator. However, the adaptive skipping of the discriminator training (freezing) did not improve the generator outputs as shown in figure \ref{Fig:cgan_freezeDisc_comparison} (see supplementary material), even though the loss functions of discriminator and generator stayed closer together. The S-parameter error converges more slowly when discriminator freezing is enabled, although it asymptotically approaches the same values. The pixel error and the classifier outputs are also overall worse when freezing is enabled.

\textbf{Blurring} In our experiments, we also employed blurring of the training set, albeit to a lesser extent. This approach aligns with the blurring used when training the forward and classifier models. By incorporating blurring during these early stages, we aim to facilitate the training of the generator network. This strategy was thought to help the generator network to gradually learn and produce more accurate and refined outputs as the training progresses. In reality, blurring did not make a significant difference in training performance as can be seen in figure \ref{Fig:cgan_blurring_comparison} ($\mathcal{S}$ parameter MSE $p=0.877$, classifier output $p=0.716$, and pixel error $p=0.883$, see supplementary material) and we therefore cannot recommend this technique in the discriminator of CGAN-based inverse design.

\begin{figure*}[t!]
    \centering
    \includegraphics[width=0.45\textwidth]{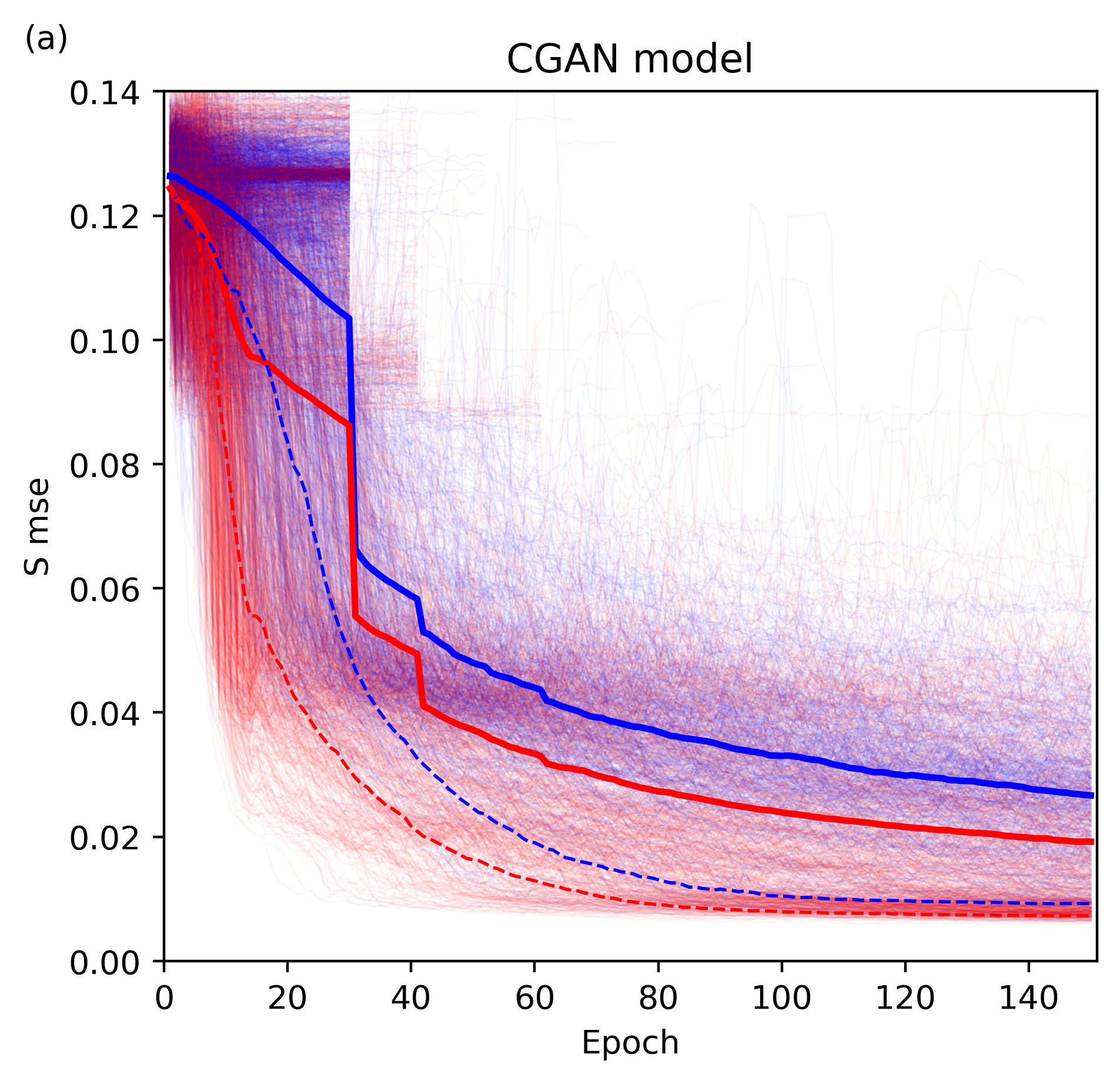}
    \includegraphics[width=0.45\textwidth]{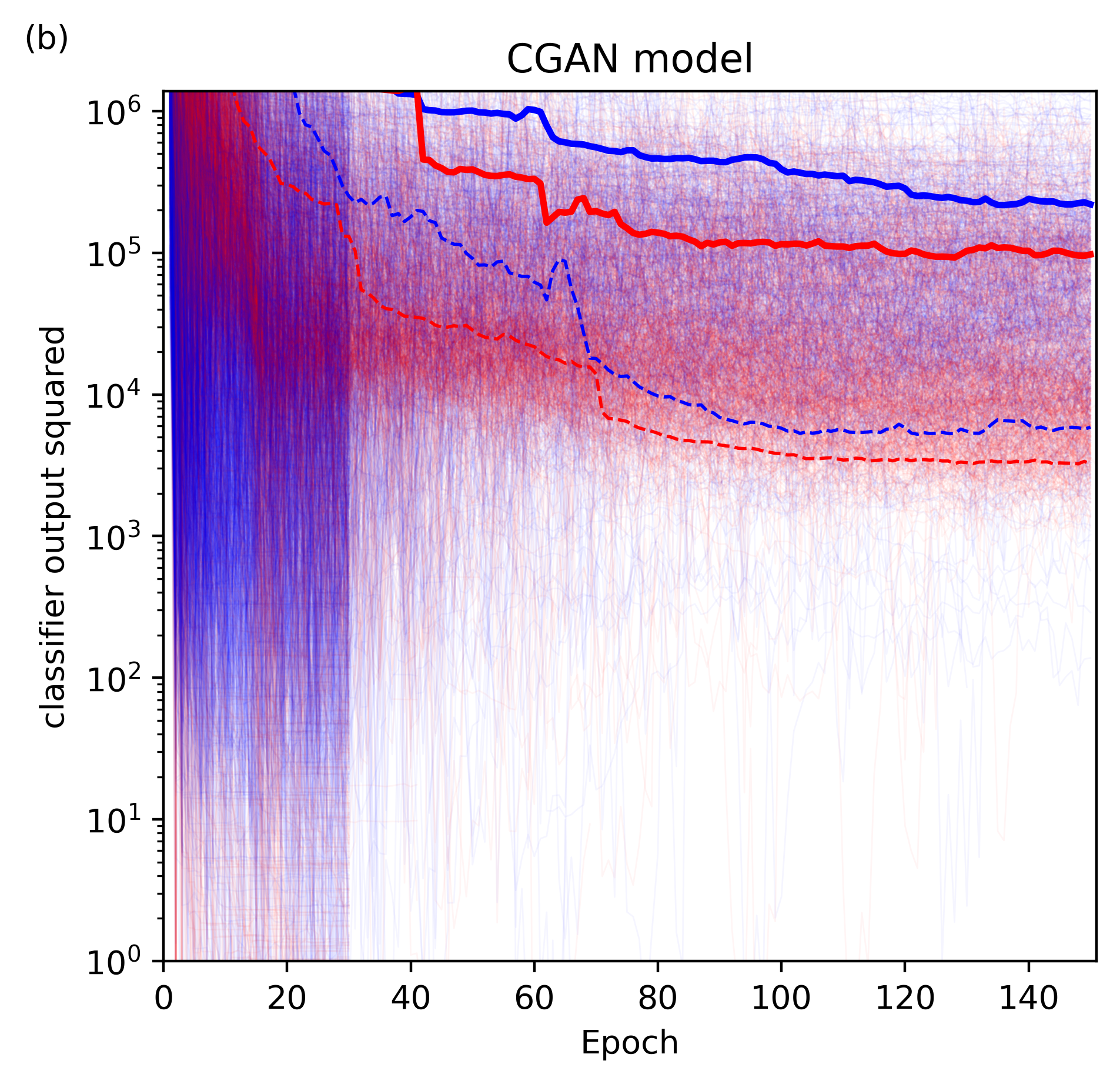}
    \includegraphics[width=0.45\textwidth,valign=m]{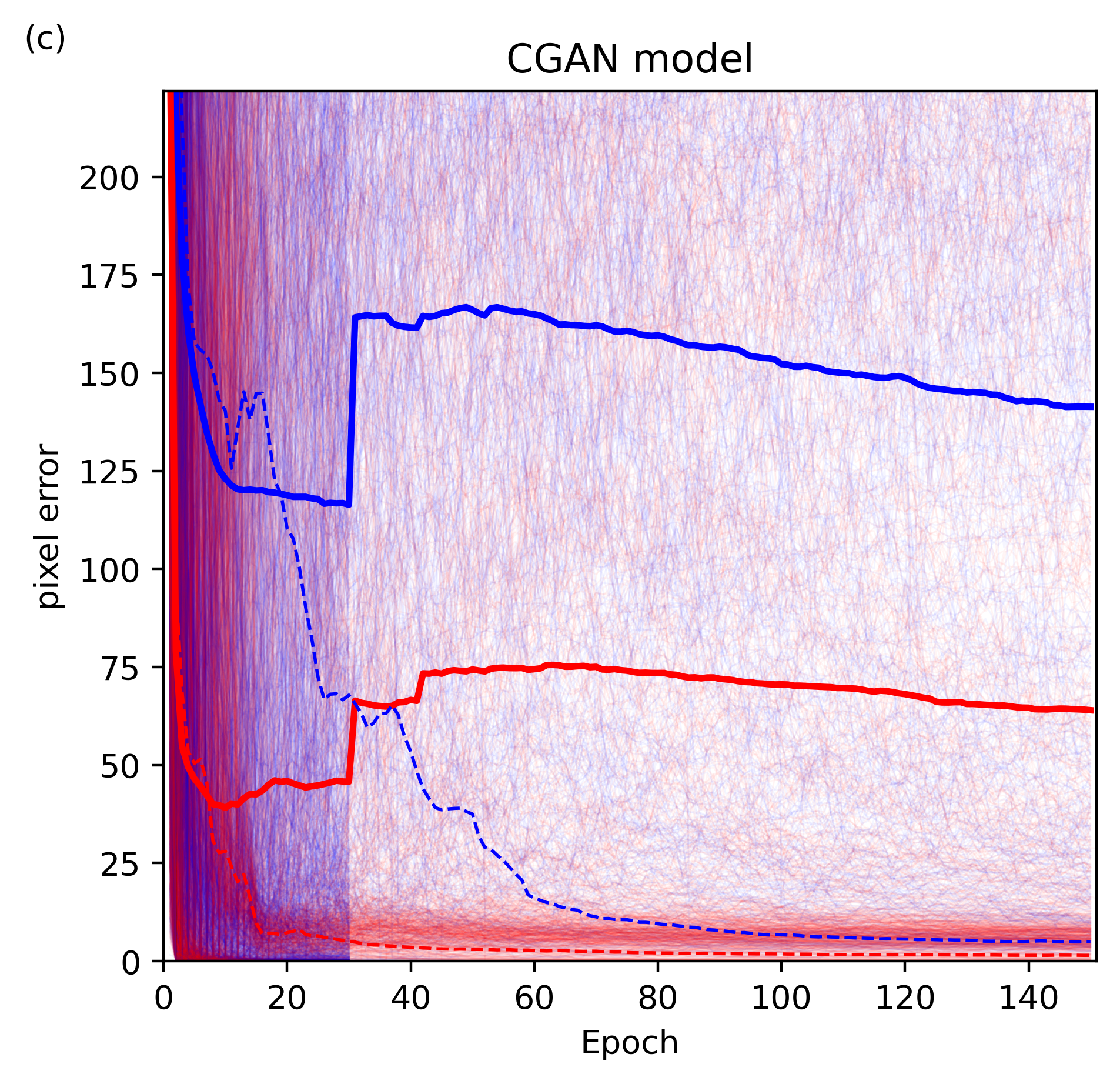}
    \includegraphics[width=0.45\textwidth,valign=m]{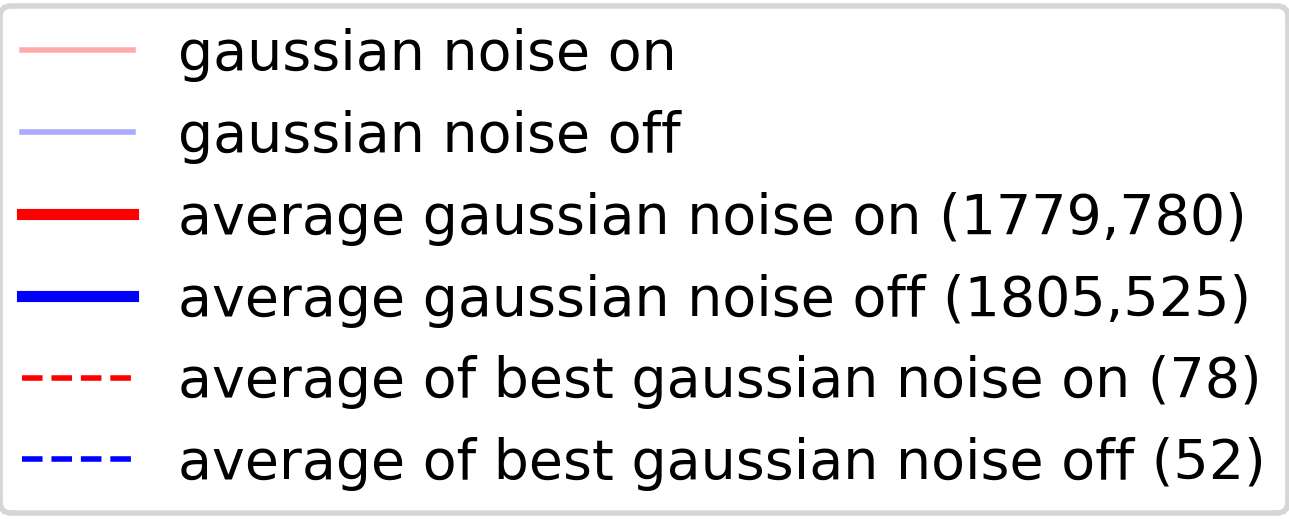}
    \caption{Comparison of the effects that Gaussian noise to the various input channels of the discriminator has on (a) $\mathcal{S}$ parameter MSE, $p<10^{-5}$, (b) classifier output, $p<10^{-5}$, and (c) pixel error, $p<10^{-5}$. In the legend, numbers after overall averages denote the total number of training instances and those reaching epoch 150. Similarly, the figure following the best training instance average indicates the number of instances used for that average.}
    \label{Fig:cgan_gaussianNoise_comparison}
\end{figure*}

\textbf{Gaussian noise} Along with the data blurring, various types of Gaussian noise~\cite{noiseneco,noise6796981,Goodfellow_2016,noiseKOTHARI1993119} were introduced to the three input channels of the discriminator during the first few epochs. A substantial amount of noise, decreasing with each epoch, was added to the input of the forward model and the classifier, both integrated into the discriminator, compelling the discriminator to rely not only on the output of the two advisor networks but also train its own CNN. This added noise proved to be crucial in improving the generator output with regards to all metrics (see figure \ref{Fig:cgan_gaussianNoise_comparison}, $\mathcal{S}$ parameter MSE, $p<10^{-5}$, (b) classifier output, $p<10^{-5}$, and (c) pixel error, $p<10^{-5}$).

\begin{figure*}[t!]
    \centering
    \includegraphics[width=0.45\textwidth]{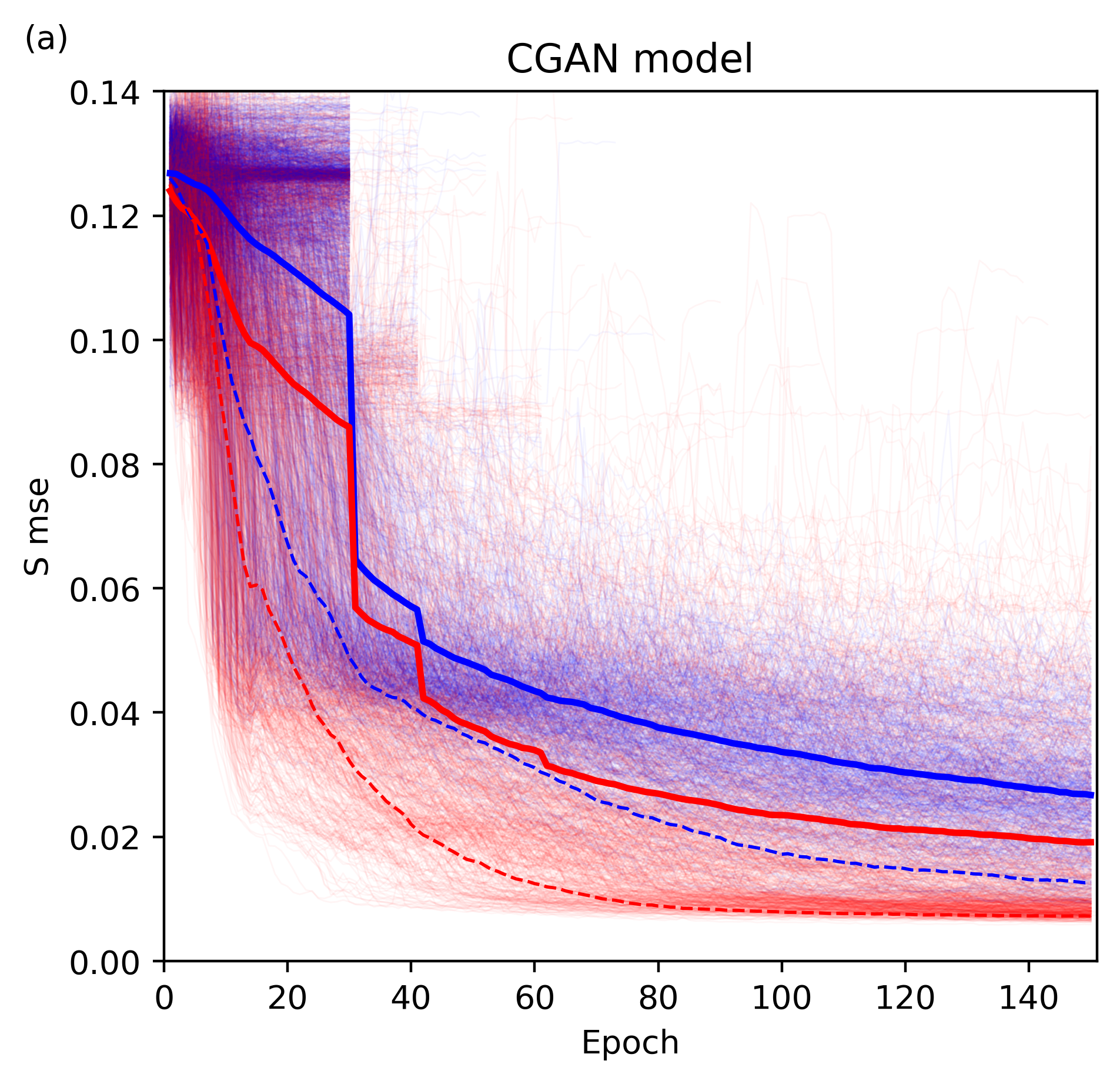}
    \includegraphics[width=0.45\textwidth]{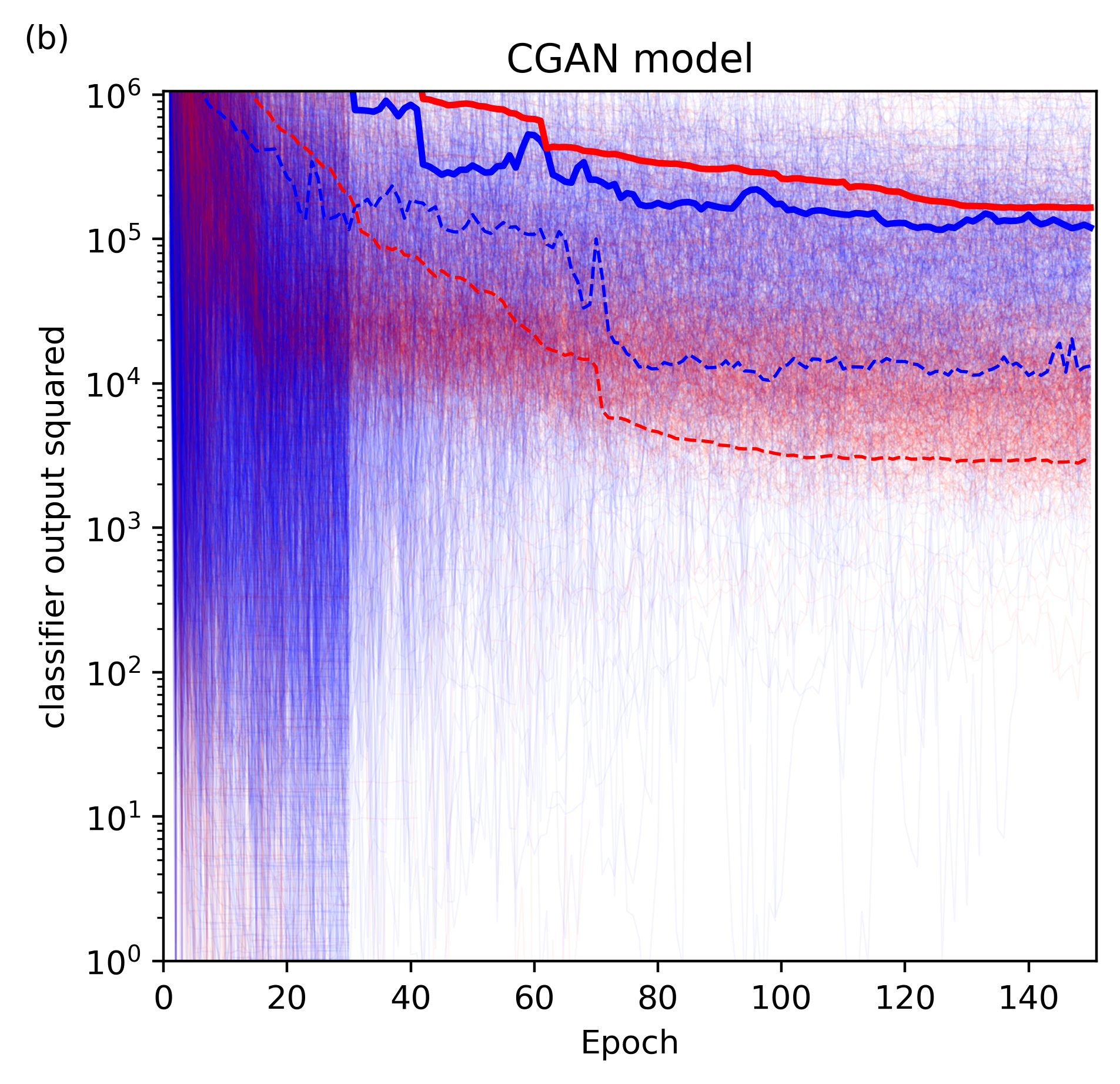}
    \includegraphics[width=0.45\textwidth,valign=m]{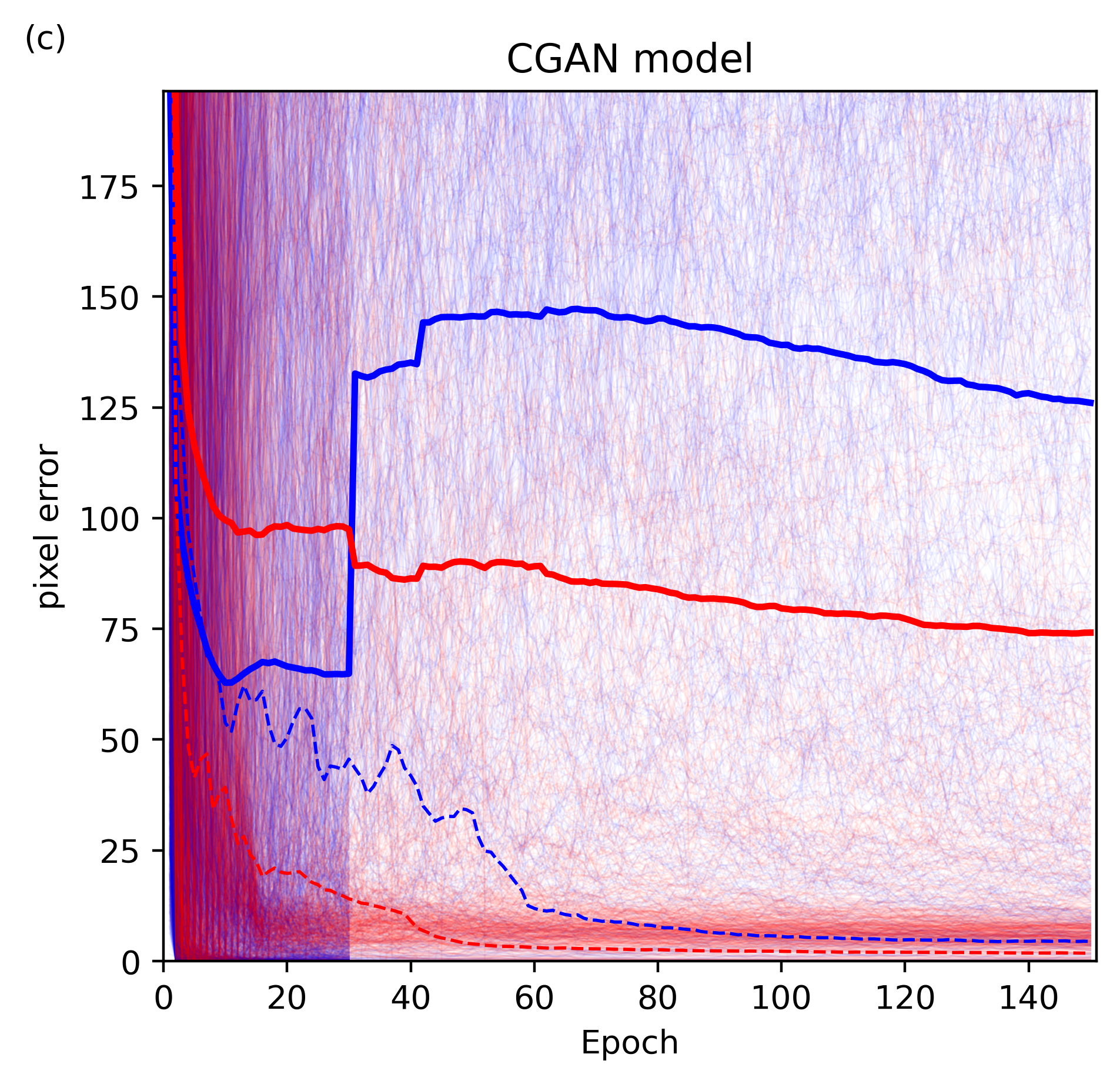}
    \includegraphics[width=0.45\textwidth,valign=m]{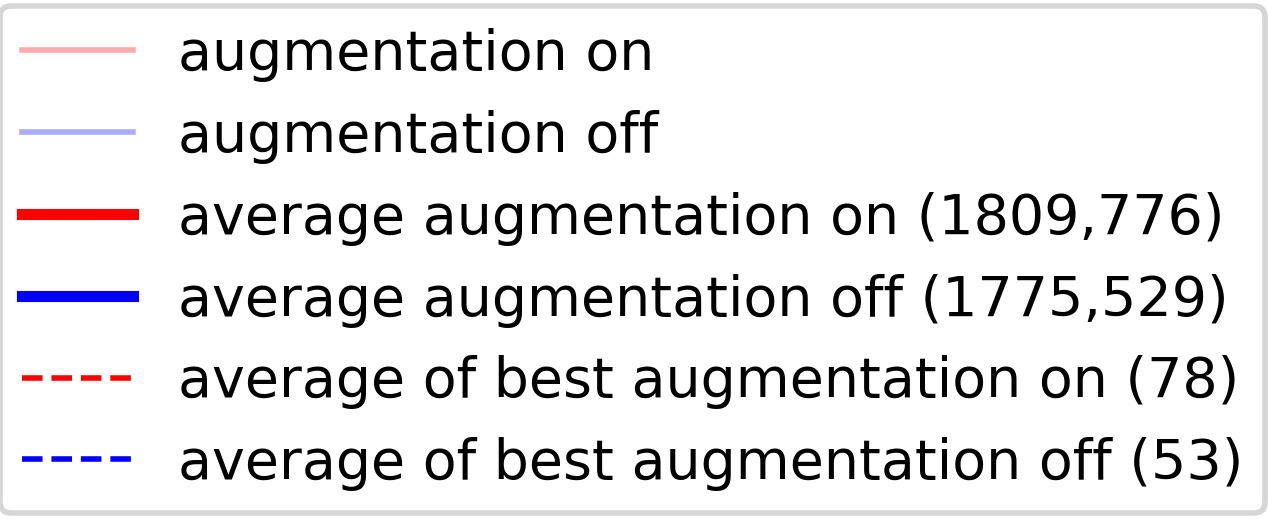}
    \caption{Comparison of the effects that augmentation of the training data and generator outputs fed to the discriminator has on (a) $\mathcal{S}$ parameter MSE, $p<10^{-5}$, (b) classifier output, $p<10^{-5}$, and (c) pixel error, $p<10^{-5}$. In the legend, numbers after overall averages denote the total number of training instances and those reaching epoch 150. Similarly, the figure following the best training instance average indicates the number of instances used for that average.}
    \label{Fig:cgan_augmentation_comparison}
\end{figure*}

\textbf{Data augmentation} We also investigated using data augmentation for all the inputs to the discriminator (both inputs coming from the generator and from the training data set). Here, we introduced the same random translation of the unit cell inputs as discussed in section~\ref{Chap:aug}). From figure \ref{Fig:cgan_augmentation_comparison}, we see that data augmentation improves the performance of the CGAN substantially. The S-parameter

 MSE and the pixel error converge faster when data augmentation is enabled $\mathcal{S}$ parameter MSE, $p<10^{-5}$, pixel error $p<10^{-5}$. For the classifier output, we see that data augmentation leads to a slightly larger average, but also that the best models perform better with data augmentation ($p<10^{-5}$). This indicates a larger variability and demonstrates the importance of training sufficient different models in order to find one of the well-performing models.

Furthermore, we automated the collection of generator inputs and outputs from all independent CGAN training instances and saved them with the current epoch as label. Random samples are taken from epochs in a defined interval around the current epoch, which are then used to train the discriminator on a diverse set of data after each epoch. The effect of this was slightly positive, though not significant (figure \ref{Fig:cgan_discCollectiveMemory_comparison} in the supplementary material; $\mathcal{S}$ parameter MSE $p=0.354$, classifier output $p=0.987$, and pixel error $p=0.223$). Additionally, if the generator error of the current batch in any epoch becomes too high, the discriminator is trained in the subsequent batch on outputs from its own previous epochs instead of the current outputs if freezing of the discriminator is enabled or with a small random change otherwise. The effects of this were also slightly positive, albeit insignificant as shown in figure \ref{Fig:cgan_discOldEpoch_comparison} (see supplementary material; $\mathcal{S}$ parameter MSE $p=0.249$, classifier output $p=0.725$, and pixel error $p=0.774$). Both alternative training schemes utilized another optimizer with a lower learning rate $\in [8\cdot 10^{-10},1\cdot 10^{-8}], \beta_1=0.333, \beta_2=0.999, \epsilon=10^{-7}$.

\subsection{Generator}

Finally, we studied how the network architecture of the generator can enhance its stability and performance. We increased the number of neurons in the fully connected layers from $2^9-2^{10}$ to $2^{10}-2^{11}$, finding that anything beyond this range caused significant instability (network architecture can be found in figure~\ref{Fig:generator}). In terms of the transverse convolutional network architectures, we conducted experiments using different configurations. The simplest architecture consisted of standard networks with batch normalizations applied between the transverse convolutional layers (called DefaultNet in figure~\ref{Fig:cgan_TransversConNets_comparison}, architecture can be found in figure~\ref{Fig:DefaultNet}). Additionally, we explored four groups of transverse convolutional layers, incorporating varying numbers of batch normalizations between the respective groups (referred to as FourNet in the figure~\ref{Fig:cgan_TransversConNets_comparison}, architecture can be found in figure~\ref{Fig:FourNet}). In the third architecture, we introduced transverse convolutional connections between the groups (HelixNet in the figure~\ref{Fig:cgan_TransversConNets_comparison}, architecture can be found in figure~\ref{Fig:HelixNet}). These different architectures are compared in figure~\ref{Fig:cgan_TransversConNets_comparison}. We find that the HelixNet performs best, although the improvements are rather small ($\mathcal{S}$ parameter MSE $p=[0.075,0.066,0.091]$, classifier output $p<10^{-5}$, and (c) pixel error $p=[0.004,0.083,0.562]$). In order to further refine the outputs, we also introduced a DenseNet block at the end of the generator's transposed convolution layers (the architecture of the DefaultNet with a DenseNet block can be found in figure~\ref{Fig:DenseNetBlock}). This addition helped to enhance the quality and accuracy of the generated outputs by a small margin, contributing to the overall success of the network (figure~\ref{Fig:cgan_denseNetBlockOut_comparison} in the supplementary material; $\mathcal{S}$ parameter MSE $p=0.437$, classifier output $p=0.37$, and pixel error $p=0.174$).

\section{Conclusion}
We have studied a number of different machine-learning techniques in CGAN-based free-form inverse design of metasurfaces. In the forward network, which maps the geometry of the optical device onto its optical properties, and in the classifier, which estimates the fabricational feasibility, we found that the use of DenseNets and data augmentation significantly improves the training of the networks. Data augmentation is particularly interesting, since the major computational cost of this form of inverse design lies in the generation of training data. Furthermore, in the context of CGAN-based inverse design, it turns out to be very advantageous to train the pre-trained expert networks on blurred data; this prepares these networks better for their task when integrated into the discriminator, where they are presented the intermediate generator outputs, which resemble more blurred data. In the CGAN, we first demonstrated that it is crucial to integrate pretrained expert networks helping the discriminator with assessing the physical properties and manufacturing feasibility---without these expert networks we could not make the training of the CGAN converge and they are therefore key to CGAN-based inverse design. Further techniques that provided most significant advantages to the training of the CGAN are data augmentation and adding noise to the inputs of the networks, both improving the performance of our CGAN for inverse design immensely. While in line with other research on CNNs, adding noise to the inputs of the pretrained expert networks is in particular important to force the discriminator to train its own CNN. Surprisingly, a number of other machine learning techniques did not improve our CGAN for inverse design (DenseNets in the discriminator, discriminator freezing) or turned out to be only marginally useful (blurring, training on other sources of generator outputs, alternative transverse convolutional architectures). This shows that the need to carefully assess what machine-learning techniques are beneficial for a specific application. While we used our network for the inverse design of optical metasurfaces, the same CGAN can also be used for other optical devices and even devices with other physical properties, as long as it is possible to simulate the devices and label them with the physical property of interest.

\begin{backmatter}
\bmsection{Funding}
We acknowledge support from Chalmers’ Excellence Initiative Nano and from the Swedish Research Council under Grant No.~2020-05284. The training data generation and ANN training were performed on resources provided by the Swedish National Infrastructure for Computing (NAISS), at the Chalmers/C3SE and KTH/PDC sites, partially funded by the Swedish Research Council under Grant No.~2022-06725.
\end{backmatter}

\tiny 

\renewcommand{\thefigure}{S\arabic{figure}}
\setcounter{figure}{0}
%\captionsetup[figure]{
%  labelfont={color=white},   % color for "Figure X"
%  textfont={color=white}     % color for the rest of the caption text
%}

%\begin{figure*}[h!]
%	\centering
%	\textcolor{white}{\caption{\label{Fig:network_forward_withdensenet}}}
%\end{figure*}
\refstepcounter{figure}\label{Fig:network_forward_withdensenet}

%\begin{figure*}[h!]
%	\centering
%	\textcolor{white}{\caption{\label{Fig:network_forward_withoutdensenet}}}
%\end{figure*}
\refstepcounter{figure}\label{Fig:network_forward_withoutdensenet}

%\begin{figure*}[h!]
%	\centering
%	\textcolor{white}{\caption{\label{Fig:network_classifier_withdensenet}}}
%\end{figure*}
\refstepcounter{figure}\label{Fig:network_classifier_withdensenet}

%\begin{figure*}[h!]
%	\centering
%	\textcolor{white}{\caption{\label{Fig:network_classifier_withoutdensenet}}}
%\end{figure*}
\refstepcounter{figure}\label{Fig:network_classifier_withoutdensenet}

%\begin{figure*}[h!]
%	\centering
%	\textcolor{white}{\caption{\label{Fig:cl_in_disc_comp}}}
%\end{figure*}
\refstepcounter{figure}\label{Fig:cl_in_disc_comp}

%\begin{figure*}[h!]
%	\centering
%	\textcolor{white}{\caption{\label{Fig:discriminator_DenseNet}}}
%\end{figure*}
\refstepcounter{figure}\label{Fig:discriminator_DenseNet}

%\begin{figure*}[h!]
%	\centering
%	\textcolor{white}{\caption{\label{Fig:discriminator_conv}}}
%\end{figure*}
\refstepcounter{figure}\label{Fig:discriminator_conv}

%\begin{figure*}[h!]
%	\centering
%	\textcolor{white}{\caption{\label{Fig:cgan_denseNet_comparison_cgan}}}
%\end{figure*}
\refstepcounter{figure}\label{Fig:cgan_denseNet_comparison_cgan}

%\begin{figure*}[h!]
%	\centering
%	\textcolor{white}{\caption{\label{Fig:cgan_freezeDisc_comparison}}}
%\end{figure*}
\refstepcounter{figure}\label{Fig:cgan_freezeDisc_comparison}

%\begin{figure*}[h!]
%	\centering
%	\textcolor{white}{\caption{\label{Fig:cgan_blurring_comparison}}}
%\end{figure*}
\refstepcounter{figure}\label{Fig:cgan_blurring_comparison}

%\begin{figure*}[h!]
%	\centering
%	\textcolor{white}{\caption{\label{Fig:cgan_discCollectiveMemory_comparison}}}
%\end{figure*}
\refstepcounter{figure}\label{Fig:cgan_discCollectiveMemory_comparison}

%\begin{figure*}[h!]
%	\centering
%	\textcolor{white}{\caption{\label{Fig:cgan_discOldEpoch_comparison}}}
%\end{figure*}
\refstepcounter{figure}\label{Fig:cgan_discOldEpoch_comparison}

%\begin{figure*}[h!]
%	\centering
%	\textcolor{white}{\caption{\label{Fig:generator}}}
%\end{figure*}
\refstepcounter{figure}\label{Fig:generator}

%\begin{figure*}[h!]
%	\centering
%	\textcolor{white}{\caption{\label{Fig:cgan_TransversConNets_comparison}}}
%\end{figure*}
\refstepcounter{figure}\label{Fig:cgan_TransversConNets_comparison}

%\begin{figure*}[h!]
%	\centering
%	\textcolor{white}{\caption{\label{Fig:DefaultNet}}}
%\end{figure*}
\refstepcounter{figure}\label{Fig:DefaultNet}

%\begin{figure*}[h!]
%	\centering
%	\textcolor{white}{\caption{\label{Fig:FourNet}}}
%\end{figure*}
\refstepcounter{figure}\label{Fig:FourNet}

%\begin{figure*}[h!]
%	\centering
%	\textcolor{white}{\caption{\label{Fig:HelixNet}}}
%\end{figure*}
\refstepcounter{figure}\label{Fig:HelixNet}

%\begin{figure*}[h!]
%	\centering
%	\textcolor{white}{\caption{\label{Fig:DenseNetBlock}}}
%\end{figure*}
\refstepcounter{figure}\label{Fig:DenseNetBlock}

%\begin{figure*}[h!]
%	\centering
%	\textcolor{white}{\caption{\label{Fig:cgan_denseNetBlockOut_comparison}}}
%\end{figure*}
\refstepcounter{figure}\label{Fig:cgan_denseNetBlockOut_comparison}

\end{document}